\newcounter{myctr}
\def\myitem{\refstepcounter{myctr}\bibfont\noindent\ifnum\themyctr>9\else\phantom{0}\fi\hangindent17pt\themyctr.\enskip}

\documentclass{IP}
\usepackage{hyperref}
\usepackage[super,sort,compress]{cite}
\usepackage{mathrsfs}
\usepackage{amsfonts}
\usepackage{amssymb}
\usepackage{amsmath}
\usepackage{epsfig}
\usepackage{makeidx}
\usepackage{setspace}
\usepackage{rotating}
\usepackage{dcolumn}
\usepackage{bm}
\usepackage{bbold}
\usepackage{graphicx}% Include figure files
\usepackage{subfigure}
\begin{document}

%%%%%%%%%%%%%%%%%%%%% Publisher's Area please ignore %%%%%%%%%%%%%%
\catchline{}{}{}{}{}
%%%%%%%%%%%%%%%%%%%%%%%%%%%%%%%%%%%%%%%%%%%%%%%%%%%%%%%%%%%%%%%%%%%

\title{Robustness of Interferometric Power to Sudden Death}

\author{Dian Zhu}

\address{Theoretical Physics Division, Chern Institute of Mathematics, Nankai University, Tianjin 300071, China}

\author{Fu-Lin Zhang}

\address{Department of Physics, School of Science, Tianjin University, Tianjin 300072, China \\
flzhang@tju.edu.cn}
%City, State ZIP/Zone, Country\\
%second\_author@group.com}

\author{Jing-Ling Chen}

\address{Theoretical Physics Division, Chern Institute of Mathematics, Nankai University, Tianjin 300071, China \\
chenjl@nankai.edu.cn}

\maketitle

%\begin{history}
%\received{Day Month Year}
%\revised{Day Month Year}
%\accepted{Day Month Year}
%\comby{(xxxxxxxxxx)}
%\end{history}

\begin{abstract}
%It is known that quantum entanglement might experience "sudden death" under Markovian environments by choosing proper initial conditions.
We study the dissipative dynamics of interferometric power as a discordlike measure in Markovian environments, such as dephasing, depolarizing, and generalized amplitude damping.
Moreover, we compare the dynamics of interferometric power and entanglement by choosing the proper initial conditions.
Our study shows that interferometric power decays asymptotically in all cases where the sudden death of entanglement appears.
Therefore, quantum metrology based on interferometric power is more robust than entanglement.
\end{abstract}


%\tableofcontents  % optional

\markboth{Authors' Names}
{Instructions for Typing Manuscripts (Paper's Title)}

\section{Introduction}	%) A SECTION HEADING

%纠缠和量子失协是两类区别于经典关联的量子关联。
Entanglement and quantum discord are two kinds of quantum correlations that differ from classical correlations.\cite{Nielsen2000,ES1935,EPR1935,PRLOllivier2001,VVedral2001}
%尽管纠缠被认为是
While entanglement is recognized as the main reason for the quantum computation advantage over classical computation, quantum discord found in separable mixed states plays a key role in quantum computing and quantum information processing.
For instance, quantum discord can be used to study applications in deterministic quantum computation with one pure qubit (DQC1),\cite{DQC1} quantum phase transmission,\cite{PRB2008,PRA80022108} and application of quantum correlations in the Grover search algorithm.\cite{HFan2010}
%Hence, a large number of discord-type measures of quantum correlation have been introduced in recent years.

%因此，近年来引入了大量量子关联的失协类度量。
In recent years, a large number of discord-type measures of quantum correlation have been introduced, such as local quantum uncertainty (LQU),\cite{PRL110240402} local quantum fisher information (LQFI),\cite{Bera2014,PRA97032326,PLA476128868} and interferometric power (IP).\cite{Girolami2014}
%尽管LQFI和IP作为失协类度量具有等价的形式，  通过定义，IP可视为最优的LQFI，因此它们作为失协类度量具有等价的形式。
%By definition, IP can be regarded as the optimal LQFI, therefore they have an equivalent form as discord-type measures.
Although LQFI and IP have an equivalent form of discord-type measures, their definitions are not the same.%\cite{Bera2014,PRA97032326,Girolami2014}
IP can be regarded as the optimal LQFI.
%介绍quantum IP作为失协类的度量,PLA476128868
The IP, proposed in the context of quantum metrology, was first introduced by Girolami et al.\cite{Girolami2014} %was initially used for quantifying the guaranteed sensitivity of the probe state in some interferometric devices.
They proved that this quantity could be a rigorous measure of discord-type quantum correlations of an arbitrary bipartite state.
As a discordlike quantum correlation measure, IP has several basic properties, such as nonnegativity, local unitary invariance, and contractivity.\cite{PRA97032326}
Furthermore, IP exists as the phenomenon of the sudden change for a two-qubit X shape states under several different kinds of quantum noises,\cite{dzhu} just like quantum discord.\cite{VVedralPRA,PRL104200401,HFan2013,PRA88suddenchange}
%[PRA80044102,PRL104200401,IJQI1350048,PRA88034304]

%介绍entanglement的sudden death现象，以及我们的研究目的和研究结果。
%各种量子性质的开放系统动力学已得到了广泛的研究
Numerous investigations recently have been focused on the dynamical behaviors of various quantum properties in open quantum systems.\cite{CJP96700,LPL16045201,CPB30030308,PS96015101,LPL19035204,Uni95,AP5352200523,AEJ8327,EPJC84670,EPJC8442}
For instance, when considering the dynamics of a two-qubit entangled state under noisy environments, disentanglement can occur in a finite time.\cite{LNP2003,HalliwellPRA2004,TYuPRL2004,SantosPRA2006,AlmeidaScience2007,AcinPRL2008,TYuPRL2006,TYuQIC2007,PRA78022322} %[PRA80024103cite9-13]which is different from the usual decoherence in asymptotic time.
%while the decoherence takes an infinite time.
%This phenomenon, which is different from the usual decoherence in asymptotic time, called "entanglement sudden death" (ESD), and it depends on 
This phenomenon, so-called "entanglement sudden death" (ESD), is different from the usual decoherence in asymptotic time, depending on the initial two-qubit state and the system-environment interactions.
%The occurrence of such a phenomenon depends on the initial two-qubit state and the system-environment interactions.
Here we investigate the dynamics of IP of a two-qubit system under several different kinds of noisy channels where the ESD can occur in the same case.
%我们的结果表明,在同样的噪声环境下，当纠缠表现出突然死亡的现象时，IP只在渐进时间内衰减。
Our results show that IP decays only in the asymptotic limit when entanglement exhibits sudden death under the same noisy environments.
That is, IP is more robust than entanglement in resisting decoherence, which implies that quantum metrology based on IP is more robust than based on entanglement.

\section{Quantification of entanglement and quantum discord}

Here, to investigate the dynamics of two-qubit states we use concurrence\cite{WoottersPRL1998} and interferometric power\cite{Girolami2014} as the quantification of entanglement and quantum discord, respectively.

The concurrence of a two-qubit state $\rho_{AB}$ can be given as $C(\rho_{AB}) = \max \{0, \lambda_1 -\lambda_2 -\lambda_3 -\lambda_4\}$, where $\lambda_i$s are the eigenvalues of the Hermitian matrix
$\sqrt{\rho_{AB} \sigma_y \otimes \sigma_y \rho_{AB}^{\ast} \sigma_y \otimes \sigma_y}$, $\rho_{AB}^{\ast}$ being the complex conjugate of $\rho_{AB}$.
The structure of density matrices we use to calculate concurrence has an $X$-shape, which means that the form of concurrence can be rewritten as 
\begin{equation} \label{Conc}
C(\rho_{AB}) = 2\max\{0,\Lambda_1,\Lambda_2\},
\end{equation}
where $\Lambda_1 = |\rho_{14}| - \sqrt{\rho_{22}\rho_{33}}$, $\Lambda_2 = |\rho_{23}| - \sqrt{\rho_{11}\rho_{44}}$.

The definition of IP for $\rho_{AB}$ with subsystem A being a qubit is\cite{PRL104200401}
\begin{equation}\label{QIP}
  \mathcal{IP}^A (\rho_{AB}) = \zeta_{min} [M],
\end{equation}
where $\zeta_{min} [M]$ is the smallest eigenvalues of the matrix $M$ with elements
\begin{equation} \label{MElement}
  M_{m,n} = \frac{1}{2} \sum_{i,l;q_i+q_l \ne 0} \frac{(q_i-q_l)^2}{q_i+q_l} \langle \psi_i| \sigma_m\otimes\mathbb{1}| \psi_l\rangle \langle \psi_l| \sigma_n\otimes\mathbb{1}| \psi_i\rangle
\end{equation}
with $\{q_i,|\psi_i\rangle\}$ being the eigenvalues and eigenvectors of $\rho_{AB}$.
By this definition, $\mathcal{IP}^A (\rho_{AB})$ can be considered as an operational and computable indicator of quantum discord.
%通过这个定义，IP可作为量子失协的一个可操作和可计算的指标。
%For the case of $\rho_{AB}$ being an X-shape density matrix, 

\section{Dynamics of concurrence and IP under noisy environments}

%在量子信息处理中，一个量子信道E可被定义称一个完全正定保迹映射，其中 是满足完备性关系的信道的Kraus算子。
%量子系统在噪声环境下的演化可以用算子和表示来描述。
%The dynamics of a quantum system under noisy environments can be described by operator-sum representation.
%When consider an initial state $\rho_{AB}$, the 

\subsection{Phase damping}

\begin{figure}[htbp]
  \centering
   \subfigure[]{
   \label{Fig1sub1}
   \includegraphics[width=0.4\linewidth]{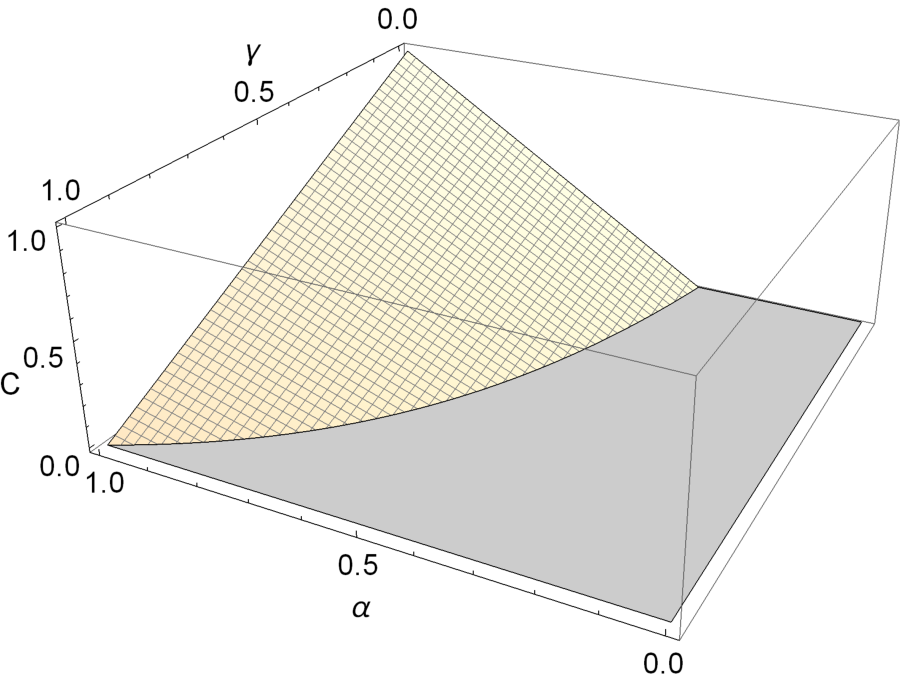}
   }
   \subfigure[]{
   \label{Fig1sub2}
  \includegraphics[width=0.4\linewidth]{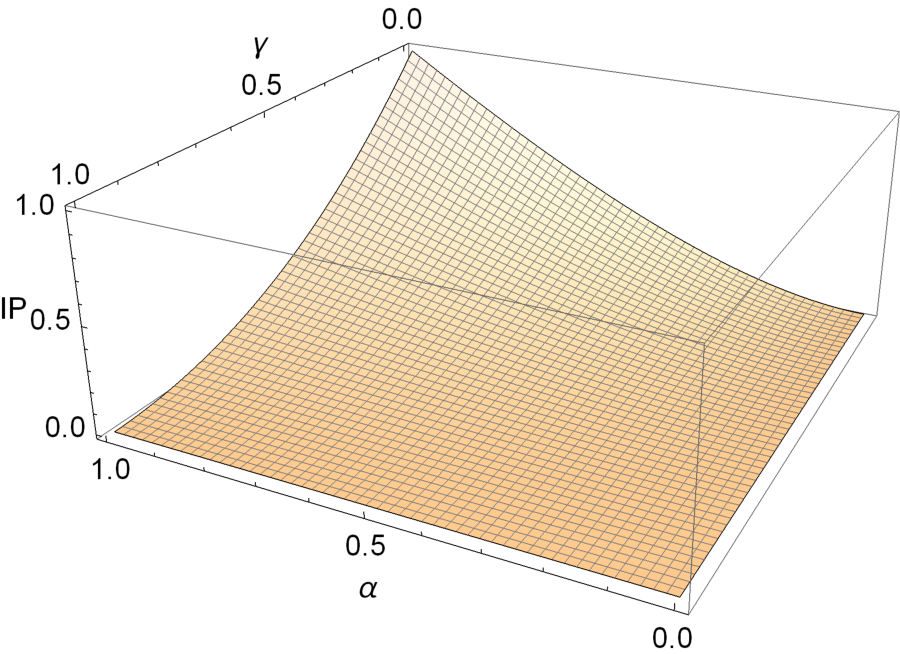}
   }
  \caption{Dynamics of (a) concurrence and (b) IP as functions of $\alpha$ and $\gamma$, under phase damping channel. }\label{Fig1}
\end{figure}

Phase damping describes the loss of information without loss of energy.
To investigate the dynamics of two-qubit entanglement and IP under dephasing noisy channel, we consider the initial state as Werner state $\rho_W(0) = (1-\alpha)\mathbb{1} / 4 + \alpha |\Psi^-\rangle\langle\Psi^-|$, where $\alpha \in [0,1]$ and $|\Psi^-\rangle = (|01\rangle -|10\rangle)/\sqrt{2}$.
The Kraus operators for a dephasing channel we considered can be given by\cite{Nielsen2000}
\begin{equation} \label{phKraus}
  E_{0} = \left( \begin{array}{cc}
             1 & 0 \\
             0 & \sqrt{1-\gamma} 
           \end{array}
           \right), \ \
  E_{1} = \left( \begin{array}{cc}
             0 & 0 \\
             0 & \sqrt{\gamma} 
           \end{array}
           \right),         
\end{equation}
where $\gamma =1 - e^{-\Gamma t}$ and $\Gamma$ denotes transversal decay rate.
During the evolution, the elements of the density matrix for output states are
\begin{equation}
  \begin{split}
     \rho_{ii}(t) &= \rho_{ii}(0), \ \ \ i=1,\dots,4 \\
     \rho_{23}(t) &= \rho_{32}(t) = \rho_{23}(0) (1-\gamma).
  \end{split}
\end{equation}
The explicit expression of concurrence for this state is $C[\rho_W(t)] = \alpha(3/2 -\gamma) -1/2 $.
It is easy to find that for any $\alpha \ne 1$, the concurrence will reach zero in a finite time, as shown in Fig.~\ref{Fig1sub1}.
On the other hand, the expression of IP for this state is $\mathcal{IP}^A[\rho_W(t)] = \min \{ \alpha^2 (2+\gamma)^2/[2(1+\alpha-\alpha\gamma)]\! +\! \alpha^2 \gamma^2/[2(1-\alpha+\alpha\gamma)], 2\alpha^2(1-\gamma)^2/(1+\alpha) \}$.
%$\mathcal{IP}(\rho) = \min \{ \frac{\alpha^2 (2+\gamma)^2}{2(1+\alpha-\alpha\gamma)} + \frac{\alpha^2 \gamma^2}{2(1-\alpha+\alpha\gamma)}, \frac{2\alpha^2(1-\gamma)^2}{1+\alpha} \}$
One can see that for any $\alpha \ne 1$, the IP of $\rho_W(t)$ vanishes only in the asymptotic limit, as shown in Fig.~\ref{Fig1sub2}.

\subsection{Generalized amplitude damping}

Generalized amplitude damping (GAD) describes the effect of dissipation to an environment at a finite temperature. %Nielsen QC & QI , 10th anniversary edition, P382
The Kraus operators of this kind of noisy channel are\cite{Nielsen2000}
\begin{equation}
  \begin{split}
     E_{0} &=\sqrt{q} \left( \begin{array}{cc}
             1 & 0 \\
             0 & \sqrt{1-\gamma} 
           \end{array}
           \right), \ \
     E_{1} =\sqrt{q} \left( \begin{array}{cc}
             0 & \sqrt{\gamma} \\
             0 & 0 
           \end{array}
           \right), \\
     E_{2} &=\sqrt{1-q} \left( \begin{array}{cc}
             \sqrt{1-\gamma} & 0 \\
             0 & 1 
           \end{array}
           \right), \ \
     E_{3} =\sqrt{1-q} \left( \begin{array}{cc}
             0 & 0 \\
             \sqrt{\gamma} & 0 
           \end{array}
           \right), 
  \end{split}
\end{equation}
where $q$ defines the probability distribution of the qubit when $t \rightarrow \infty$ and $\gamma$ is defined as in Eq.~\ref{phKraus}.

\begin{figure}[htbp]
  \centering
   \subfigure[]{
   \label{Fig2sub1}
   \includegraphics[width=0.4\linewidth]{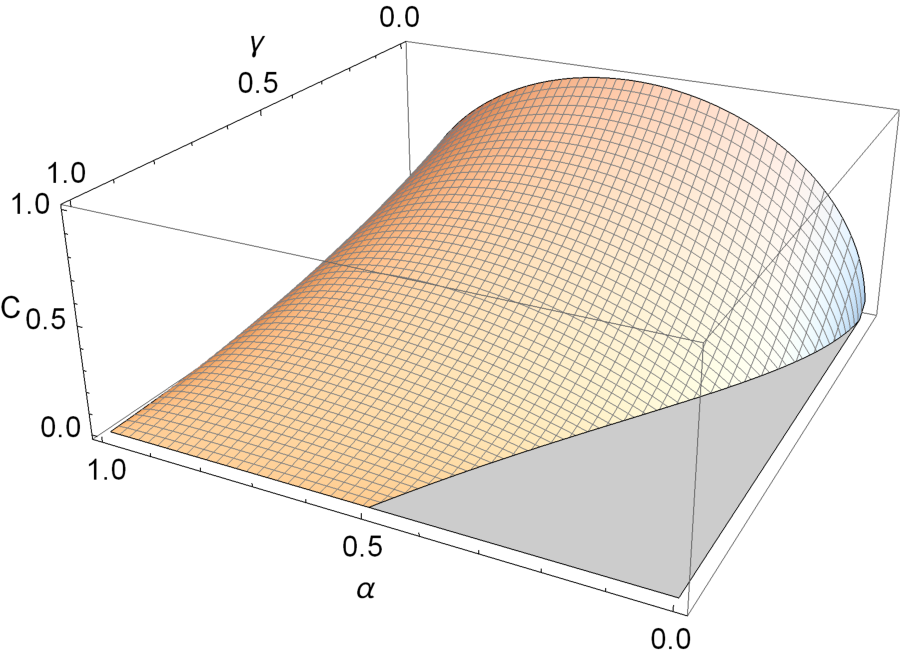}
   }
   \subfigure[]{
   \label{Fig2sub2}
  \includegraphics[width=0.4\linewidth]{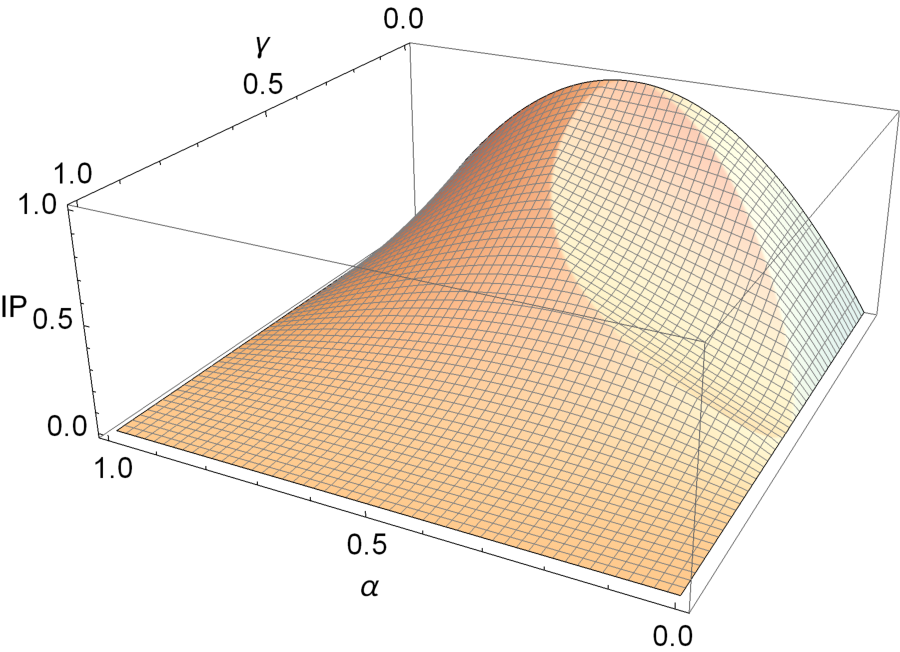}
   } 
   \subfigure[]{
   \label{Fig2sub3}
   \includegraphics[width=0.4\linewidth]{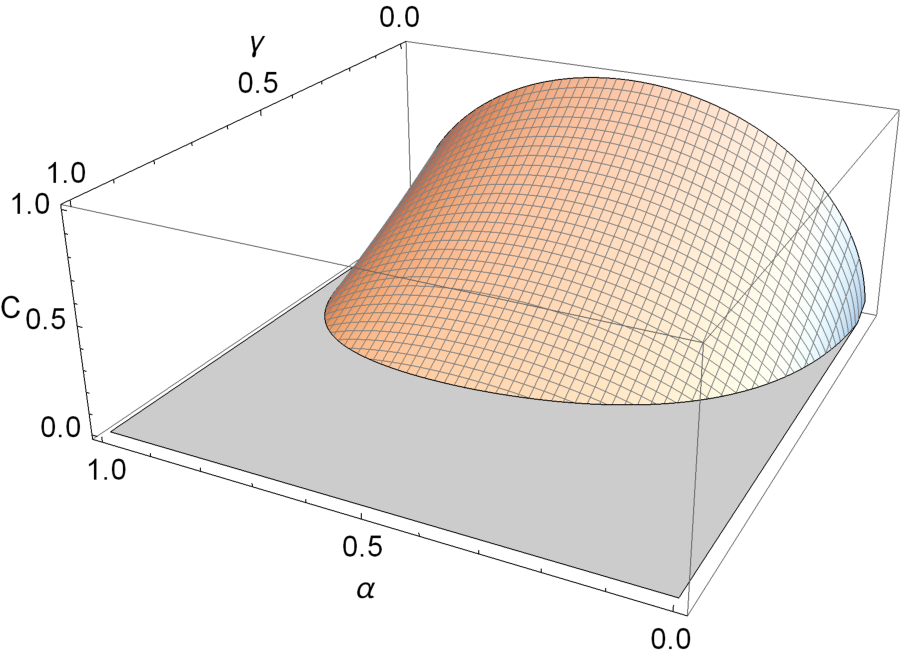}
   }
   \subfigure[]{
   \label{Fig2sub4}
  \includegraphics[width=0.4\linewidth]{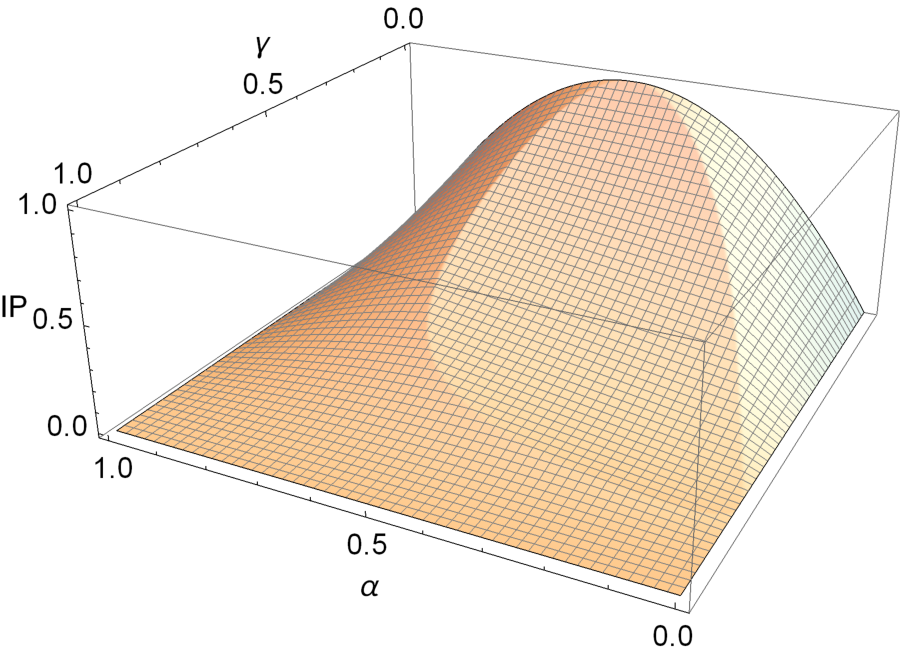}
   }
  \caption{Dynamics of (a) and (c) concurrence and (b) and (d) IP as functions of $\alpha$ and $\gamma$, under generalized amplitude damping channel. }\label{Fig2}
\end{figure}

Here we consider the initial state as $\rho(0) = |\Phi\rangle \langle\Phi|$ with 
\begin{equation}\label{Schmidt}
  |\Phi\rangle = \sqrt{1-\alpha} |00\rangle + \sqrt{\alpha} |11\rangle, \ \ \ \ \alpha \in [0,1].
\end{equation}
During the evolution, the output density matrix under the GAD channel can be obtained,
\begin{equation}
  \begin{split}
     \rho_{11}(t) =(1-\alpha) \{1 &-\gamma[2(1-q) - \gamma(1-2q)] \} + \gamma^2 q^2, \\
     \rho_{22}(t) = \rho_{33}(t) =\gamma[ & (1-\alpha)(1-2q)(1-\gamma) + q(1-\gamma q)], \\
     \rho_{44}(t) &= 1-\rho_{11}(t) - 2\rho_{22}(t), \\
     \rho_{14}(t) = &\rho_{41}(t) = (1-\gamma) \sqrt{\alpha (1-\alpha)}.
  \end{split}
\end{equation}
Using the formula (\ref{Conc}-\ref{MElement}), the concurrence and the IP of this output state can be obtained as
$C_{GAD}[\rho(t)] = 2\max\{0, \Lambda_1(t)\}$ and $\mathcal{IP}^{A}[\rho(t)] \!=\! \min \{M_{11}, M_{22}, M_{33} \}$, respectively.
%\begin{equation}
%\begin{split}
%C_{GAD}[\rho(t)] \!=\! 2\max\{0,(1&\!-\!\gamma) \sqrt{\alpha(1\!-\!\alpha)} \!-\!2\gamma [(1\!-\!\alpha)(1\!-\!2q)(1\!-\!\gamma)+q(1\!-\!\gamma q)]\},  \\
%    &\mathcal{IP}^{A}[\rho(t)] \!=\! \min \{M_{11}, M_{22}, M_{33} \},
%\end{split}
%\end{equation}
Here
\begin{equation}
\begin{split}
\Lambda_1(t)= &(1 -\gamma) \sqrt{\alpha(1 - \alpha)} - \gamma [(1 - \alpha)(1 - 2q)(1 - \gamma)+q(1 - \gamma q)], \\
M_{11} &= \frac{(\lambda_1-\lambda_3)^2 (\lambda_2+\lambda_3) +a^2(\lambda_2-\lambda_3)^2 (\lambda_1+\lambda_3)}{(\lambda_1+\lambda_3)(\lambda_2+\lambda_3)(1+a^2)} \\
   + &\frac{(\lambda_1-\lambda_4)^2 (\lambda_2+\lambda_4) + b^2(\lambda_2-\lambda_4)^2 (\lambda_1+\lambda_4)}{(\lambda_1+\lambda_4)(\lambda_2+\lambda_4)(1+b^2)} = M_{22}, \\
   &M_{33} = \frac{(\lambda_3-\lambda_4)^2}{\lambda_3+\lambda_4} \cdot \frac{(ab-1)^2}{(1+a^2)(1+b^2)},
\end{split}
\end{equation}
and $\lambda_i$ are the eigenvalues of $\rho(t)$, 
%\begin{equation}
%\begin{split}
$a =\frac{\rho_{11}(t) - \rho_{44}(t) - \sqrt{(\rho_{11}(t) - \rho_{44}(t))^2 + 4\rho_{14}^2(t)}}{2\rho_{14}(t)}$, %\\
$b =\frac{\rho_{11}(t) - \rho_{44}(t) + \sqrt{(\rho_{11}(t) - \rho_{44}(t))^2 + 4\rho_{14}^2(t)}}{2\rho_{14}(t)}$.
%\end{split}
%\end{equation}

For $q=1$, we have $\Lambda_1(t) = 2 (1-\gamma)[\sqrt{\alpha(1-\alpha)}-\alpha\gamma] >0$ for all $t$, whenever $\alpha < 1/2$ (see Fig.~\ref{Fig2sub1}).
%This implies that the ESD appears with the initial condition $\alpha >1/2$.
On the other hand, the concurrence vanishes ($\Lambda_1(t\rightarrow\infty) <0$) for all $\alpha$ when $q \ne 1$ (see Fig.~\ref{Fig2sub3}).
However, the IP exhibits different behavior compared with the concurrence.
In Figs.~\ref{Fig2sub2} and \ref{Fig2sub4}, it is easy to see that the IP decays monotonically and vanishes asymptotically in both situations ($q=1$ and $q=2/3$),
and the IP equals zero all the time for $\alpha =0$ and $\alpha =1$ (pure separable states).
In addition, an interesting result is that the dynamics of IP exhibit sudden change under the GAD channel while the quantum discord decays smoothly.\cite{PRA80024103}

\subsection{Depolarizing}

\begin{figure}[htbp]
  \centering
   \subfigure[]{
   \label{Fig3sub1}
   \includegraphics[width=0.4\linewidth]{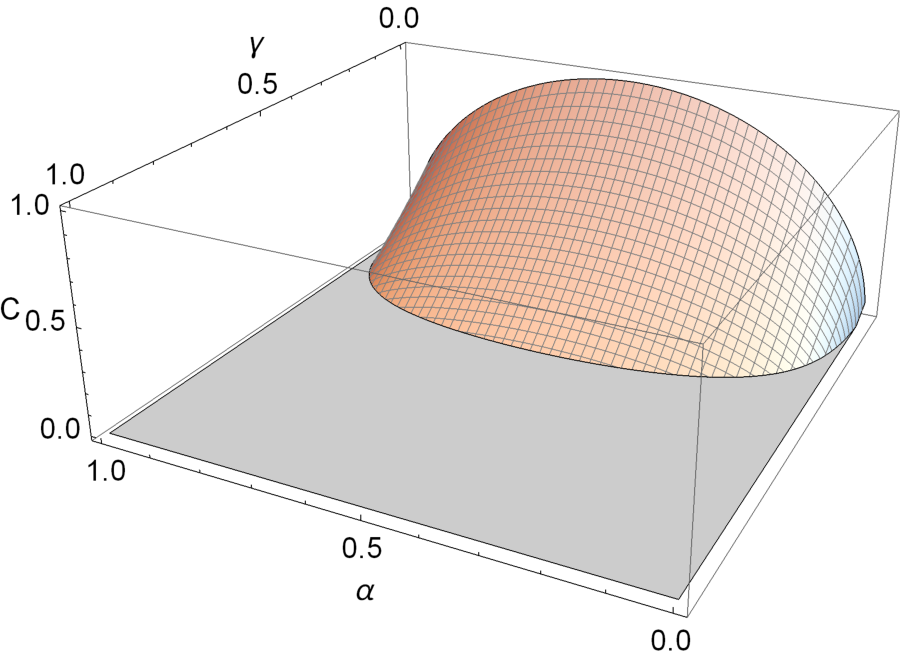}
   }
   \subfigure[]{
   \label{Fig3sub2}
  \includegraphics[width=0.4\linewidth]{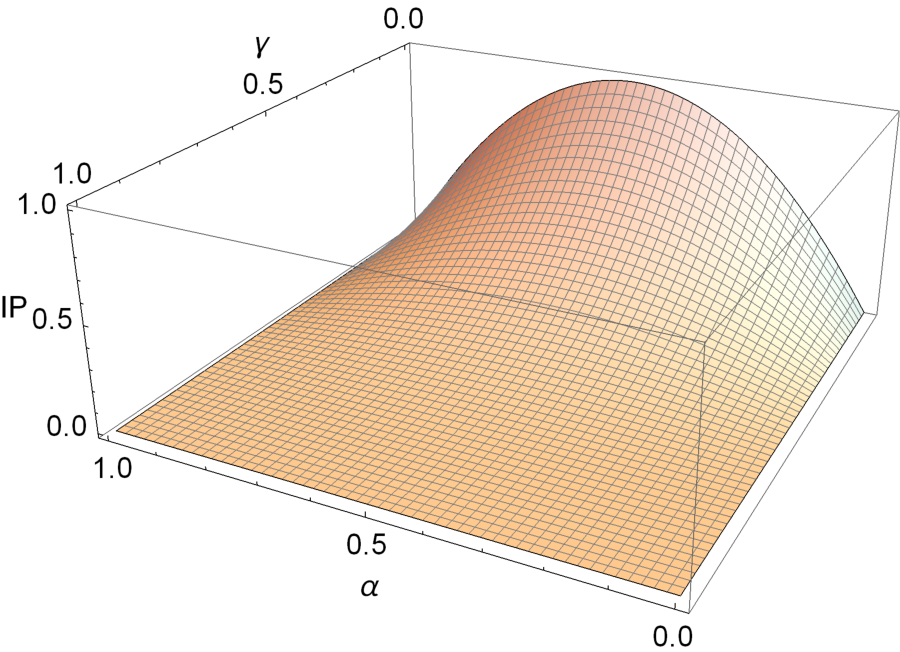}
   }
  \caption{Dynamics of (a) concurrence and (c) IP as functions of $\alpha$ and $\gamma$, under depolarizing noise channel. }\label{Fig3}
\end{figure}

Depolarizing channel describes a process that takes a state into a completely mixed state $\mathbb{1}/2$ with probability $\gamma$ and the state being left untouched with probability $1-\gamma$.
The Kraus operators of this channel are $E_0 = \sqrt{1-3\gamma/4}\mathbb{1}$, $E_1 = \sqrt{\gamma/4} \sigma_1$, $E_2 = \sqrt{\gamma/4} \sigma_2$, $E_3 = \sqrt{\gamma/4} \sigma_3$, where $\gamma$ is defined as in Eq.~\ref{phKraus}.\cite{Nielsen2000}
Similarly, we assume the initial condition as Eq.~(\ref{Schmidt}), and then we have the output density matrix elements:
\begin{equation}  %\label{}
  \begin{split}
\rho_{11}(t) &= (1-\alpha) (1-\gamma) +\gamma^2 /4, \\
\rho_{22}(t) &= \rho_{33}(t) = (2 - \gamma)\gamma/4, \\
\rho_{44}(t) &= 1 - \rho_{11}(t) - 2\rho_{22}(t), \\
\rho_{14}(t) &= \rho_{41}(t) = \sqrt{\alpha(1-\alpha)} (1-\gamma)^2.
  \end{split}
\end{equation}
By Eq.~(\ref{Conc}), the explicit expression of concurrence for this state is $C(\rho) =2\max\{0, \alpha (1-\gamma) - (2 - \gamma)\gamma/4\}$.
It is easy to find that $\alpha (1-\gamma) - (2 - \gamma)\gamma/4 < 0$ when $t \rightarrow \infty$ for all $\alpha$ (see Fig.~\ref{Fig3sub1}).
On the other hand, IP does not disappear in a finite time (see Fig.~\ref{Fig3sub2}).
Here the expression of IP can be obtained as the same procedure above.

\subsection{Dephasing and amplitude damping}

\begin{figure}
  \centering
   \subfigure[]{
   \label{Fig4sub1}
   \includegraphics[width=0.4\linewidth]{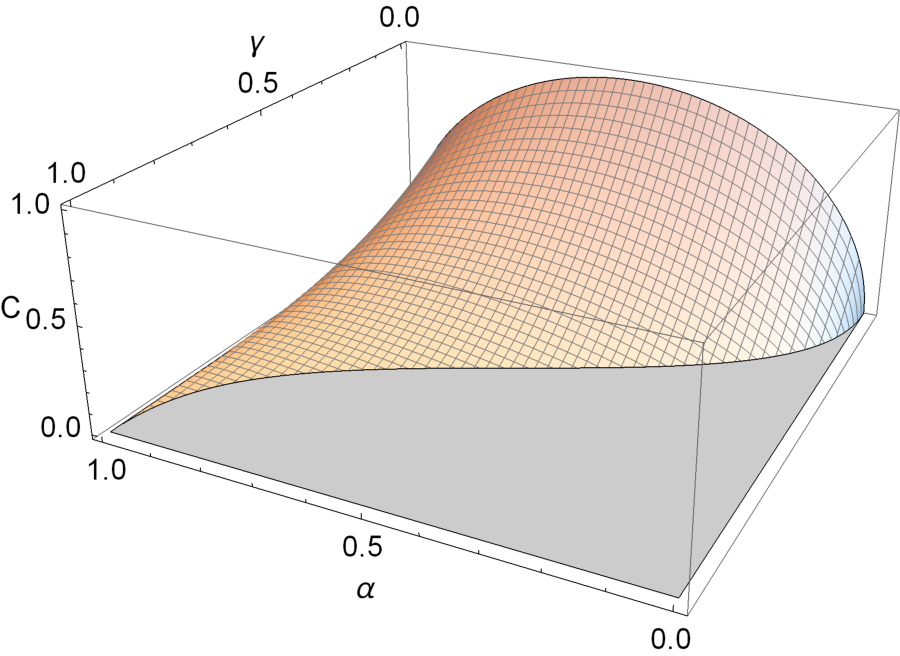}
   }
   \subfigure[]{
   \label{Fig4sub2}
  \includegraphics[width=0.4\linewidth]{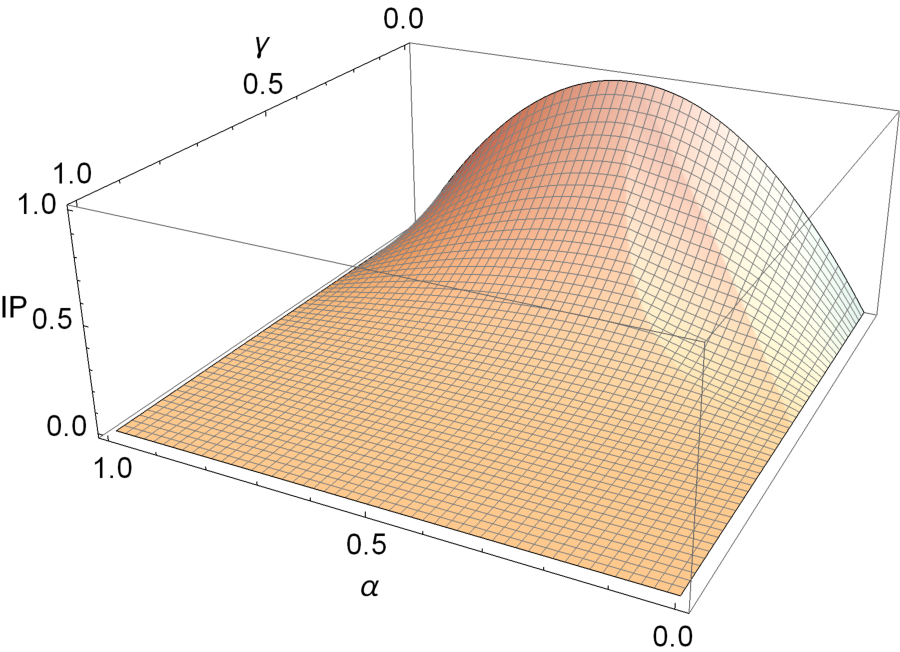}
   }
  \caption{Dynamics of (a) concurrence and (b) IP as functions of $\alpha$ and $\gamma$, under dephasing and generalized amplitude damping ($q=1$) channels acted simultaneously.}\label{Fig4}
\end{figure}

%这里我们考虑两种不同的噪声（广义振幅阻尼和相位阻尼）各自独立作用在二量子比特系统的情况。
Now we consider that two distinct reservoirs (GAD and phase damping) interact with a two-qubit system individually.
Given the density matrix of the initial state, the density matrix elements decay at the sum of the amplitude and phase rates, respectively.
% two-qubit system interact individully with two distinct reservoirs: phase damping and amplitude damping.
%Because the density matrix elements decay respectively at the sum of the amplitude and phase rates, 
%正如图中所显示的那样，单独的GAD （for q=1）并不能使得纠缠突然死亡（对alpha<1/2）。
For simplicity, we assume equal decay rates ($\Gamma$) for both channels, $q=1$ and $\gamma = 1-e^{-\Gamma t}$.
For the case of the initial condition as Eq.~(\ref{Schmidt}), neither amplitude (for $\alpha < 1/2$) nor dephasing channel (for all $\alpha$) alone can induce the sudden death of entanglement. 
%然而，当两种噪声同时应用时，纠缠的演化会截然不同。
However, entanglement dynamics are strikingly different when these two kinds of noises are applied together.
The concurrence here appears ESD for almost all $\alpha$, as shown in Fig.~\ref{Fig4sub1}.
This phenomenon of nonadditivity of decoherence while the relaxation rate is additive was first indicated by Yu and Eberly.\cite{TYuPRL2006}
%尽管如此，在相同的退相干信道上IP仍然是在渐进时间内衰减
However, the IP decays asymptotically and smoothly at the same decoherence channel (see Fig.~\ref{Fig4sub2}), which implies that the additivity of the decoherence channel is still valid for IP.

\section{Conclusion}

%总之，我们研究了二量子比特系统与独立的噪声信道耦合的IP的耗散动力学。
In summary, we investigate the dissipative dynamics of IP for a two-qubit system coupling to independent noisy channels.
%当考虑初始条件使得系统的纠缠发生突然死亡的现象时，其IP只在渐进时间内消失。
We observed that by choosing the initial conditions properly, the entanglement of the quantum system suddenly disappears while the IP vanishes only in an asymptotic manner.
This observation indicates that IP is more robust than entanglement and can be considered a better measure of the quantum resources used for some quantum information and computation processes.
%However, we observed that the dynamics of IP exhibit sudden change under the GAD channel only, but it decays smoothly when GAD and dephasing damping are applied together.
%This phenomenon might be deserved endeavor in the further investigation for the property of the IP.
%However, the existence of sudden death of IP in non-Markovian environments is unknown

\section*{Acknowledgments}

F. L. Z. was supported by the National Natural Science Foundations of China (Grants Nos. 11675119). 
J. L. C was supported by the National Natural Science Foundations of China (Grants Nos. 12275136 and 12075001).
D. Z. was supported by the Nankai Zhide Foundations.

\end{document}